\documentclass[aps,prl,twocolumn,showpacs,superscriptaddress]{revtex4}%
\usepackage{amsmath}
\usepackage{amsfonts}
\usepackage{amssymb}
\usepackage{graphicx}
\usepackage{bm}

\newcommand{\fe}{Fe$_{1/4}$TaS$_{2}$}

\begin{document}

\title{Local-moment ferromagnetism and unusual magnetic domains in Fe$_{1/4}$TaS$_{2}$ crystals}

\author{M.~D.~Vannette}
\affiliation{Ames Laboratory and Department of Physics \& Astronomy, Iowa State University, Ames, IA 50011}
\author{S.~Yeninas}
\affiliation{Ames Laboratory and Department of Physics \& Astronomy, Iowa State University, Ames, IA 50011}
\author{E.~Morosan}
\affiliation{Department of Physics \& Astronomy, Rice University, Houston, TX 77005}
\author{R.~J.~Cava}
\affiliation{Department of Chemistry, Princeton University, Princeton, NJ 08544}
\author{R.~Prozorov}
\email[Corresponding author: ]{prozorov@ameslab.gov}
\affiliation{Ames Laboratory and Department of Physics \& Astronomy, Iowa State University, Ames, IA 50011}

\date{16 May 2009}

\begin{abstract}
Single crystals of \fe~have been studied by using magneto-optical (MO) imaging and radio-frequency (rf) magnetic susceptibility, $\chi$. Real time MO images reveal unusual, slow dynamics of dendritic domain formation, the details of which are strongly dependent upon magnetic and thermal history.  Measurements of $\chi(T)$ show well-defined, local moment ferromagnetic transition at $T\approx 155$ K as well as thermal hysteresis for 50 K$<T<$60 K.  This temperature range corresponds to the domain formation temperature as determined by MO. Together these observations provide strong evidence for local moment ferromagnetism in \fe~crystals with large, temperature dependent magnetic anisotropy.

\end{abstract}


\maketitle

\begin{figure*}[tb]
\includegraphics[width=18cm]{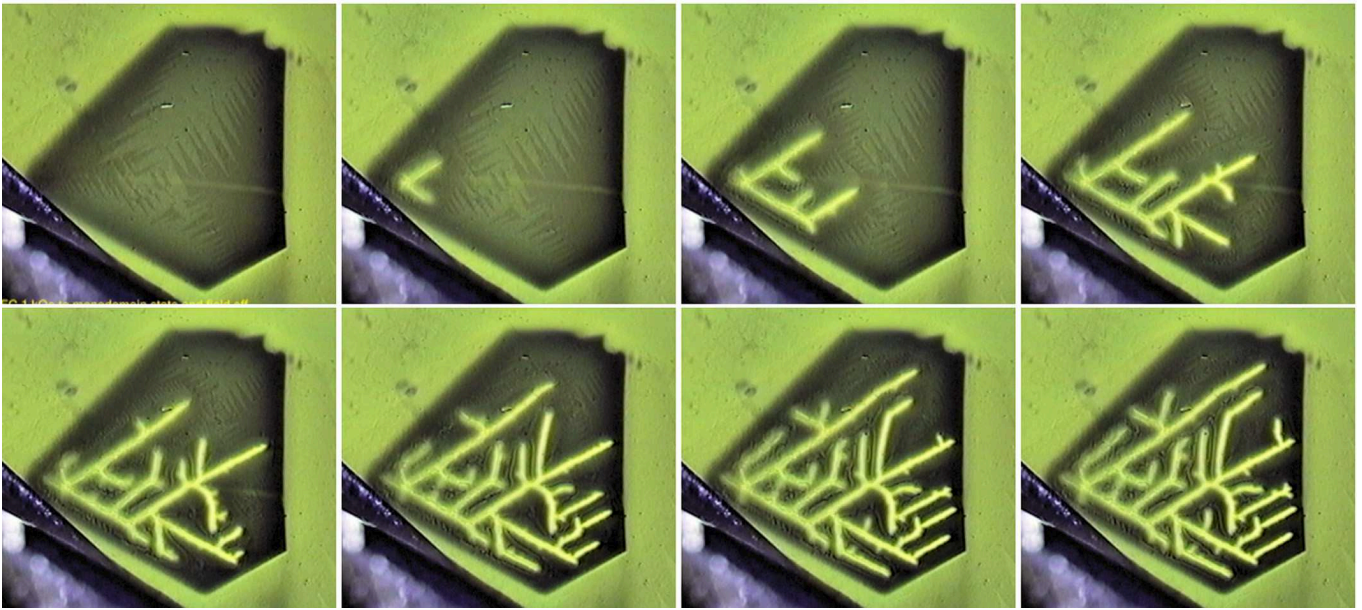}
\caption{(Color online) Applying positive magnetic field at $T = 5$ K after cooling in negative $H = - 1000$ Oe and turning field off. Top row, left to right: $H=0,\, 500,\,600,\,700$ Oe. Bottom row, left to right: $H=850,\, 1000,\,1200,\,1500$ Oe.}
\label{remag}
\end{figure*}

\section{\label{intro}Introduction}
The layered structure of the transition metal dichalcogenide TaS$_{2}$ permits the intercalation of other metal ions, nominally changing the formula to $M_{x}$TaS$_{2}$.  If $M$ is a 3$d$-transition metal, under certain circumstances long range magnetic order may occur.  The details of the order strongly depend on the type, $M$, and amount, $x$, of intercalated ion.  If $M=$ Fe, long range magnetic order is observed for $0.2\leq x\leq 0.34$.  Further, for $x=$1/4 and 1/3 ordered superlattice form with basal plane axes a=2a$_0$ and a=$\sqrt{3}$a$_0$, respectively \cite{fe-eibschutz-1981}.  The magnetic ordering temperature is non-monotonic with doping concentration, reaching a maximum at approximately 155 K for $x=1/4$ \cite{fe-eibschutz-1981}.  Band structure calculations for $x=1/3$ suggest a large imbalance between the two spin directions for the conduction electrons contributed by the iron \cite{fe-dijkstra-1989}, which may hint at a band component of the ferromagnetism.  For the $x=1/4$ concentration, however, the measured ratio of the paramagnetic moment ($\approx$4.9 $\mu_{B}$/Fe) to saturation ($\approx$4 $\mu_{B}$/Fe) moments is only slightly larger than 1 \cite{fe-morosan-2007}.  This suggests that for $x=1/4$ the system is more localized than the case where $x=1/3$ \cite{rhodes-1963}.

Recent work on single crystals of \fe~has shown extremely sharp switching fields in the $M-H$ loops \cite{fe-morosan-2007} and anomalous Hall effect and magnetoresistance associated with this sharp switching \cite{fe-checkelsky-2008}.  The time dependence of the switching was reported in Ref.~\onlinecite{fe-morosan-2007}.  There it was found that the switching speed increased almost linearly with temperature in the range $5\leq T\leq18$ K.  The time for magnetization reversal fell in the range of 0.1-0.8 ms. Several effects govern the rate of magnetization reversal in materials; the dynamics of domain evolution and eddy current dissipation in the sample due to the changing magnetic induction being of primary concern. 

In this paper we study the magnetic domain structure and formation in single crystals of \fe~by utilizing direct magneto-optical (MO) visualization based on the Faraday effect in transparent ferrimagnetic iron-garnet indicators. The magneto-optic study reveals unusual slow domain formation and sensitivity of the domain structure on the magnetic and thermal history. In addition, the radio frequency (rf) magnetic susceptibility was measured at 27 MHz via a sensitive tunnel diode resonator (TDR) technique. This technique has been shown to be a sensitive tool to distinguish between local--moment and itinerant ferromagnetism \cite{vannette-2008,Vannette2008,Vannette2008a}. The results from \fe~strongly suggest local moment ferromagnetic phase transition at 155 K. In addition, rf susceptibility shows thermal hysteresis at approximately the same temperatures that domains are seen to form in MO data.

\section{\label{expr}Experimental}
Single crystals of \fe~were grown via iodine vapor transport and characterized as described in Ref.~\onlinecite{fe-morosan-2007}.  Samples were platelets with the crystallographic $c-$axis perpendicular to the plane of the plate.  This axis is also the magnetic easy axis in these materials. Two different samples have been measured  and they produced almost identical results.

Magneto-optical (MO) imaging was performed in a flow-type optical $^{4}$He cryostat using Faraday rotation of polarized light in an indicator based on Bi-doped iron-garnet films with in-plane magnetization \cite{Prozorov2006g}. The indicator is placed directly on top of a sample with two parallel surfaces in the $x$-$y$ - plane. The light propagates through the indicator along the $z-$ axis and reflects back from a mirror sputtered at the bottom of the indicator; the sample itself is not seen. The $z-$ component of local magnetization in the indicator is proportional to the $z-$ component of the magnetic induction on the sample surface. Optical contrast shows the distribution of local magnetization in the sample, with brighter areas corresponding to larger magnetic induction values. Image color provides a distinction between positive and negative field (yellow vs. green, see electronic version of the paper with color figures).

Radio-frequency magnetic susceptibility, $\chi_{rf}$, has been measured using a tunnel-diode resonator technique (TDR)\cite{vandegrift-1975}. This technique detects changes in magnetic susceptibility through induced frequency shifts in an $LC$ circuit.   For this work, the nominal resonant frequency was 27 MHz. The tank circuit is self-resonating with losses compensated by a properly biased tunnel diode.  The sample is mounted on a sapphire rod and inserted in the coil.  The coil and sample are thermally isolated from one another. Changes in the sample's magnetic susceptibility induce changes in the resonant frequency and it is this frequency shift that is measured.  It is straightforward to show that $\Delta f/f_{0}=-G\chi_{rf}$ where $f_{0}$ is the empty coil resonant frequency, and $\Delta f$ is the change in the resonance induced by placing the sample in the coil \cite{clover-1970}.  $G$ is a constant factor that depends on sample shape and the ratio of the sample to coil volumes.  Determining the value of $G$ is difficult leading to complications when making precise statements regarding the absolute value of $\chi_{rf}$.  However, \textit{changes} in the susceptibility may be probed with extreme sensitivity.  Precise control of the temperature of the circuit components and careful circuit design together allow for a resonance stability on the order of 0.05-0.5 Hz over several hours, a resolution of 50 parts per billion.  With a typical sample size of 1 mm$^3$ this translates to a sensitivity to changes in $\chi_{rf}$ on the order of $10^{-7}$ in absolute cgs units. The excitation magnetic field produced by the tank coil is very small, $<10$ mOe, allowing for a weakly perturbative test of the sample. The TDR technique has been shown to be well suited to the study of penetration depth in superconductors \cite{prozorov-2006-2} and various ferromagnetic properties, particularly distinguishing between local-moment and itinerant ferromagnets \cite{vannette-2008}.

\begin{figure*}[t]
\includegraphics[width=18cm]{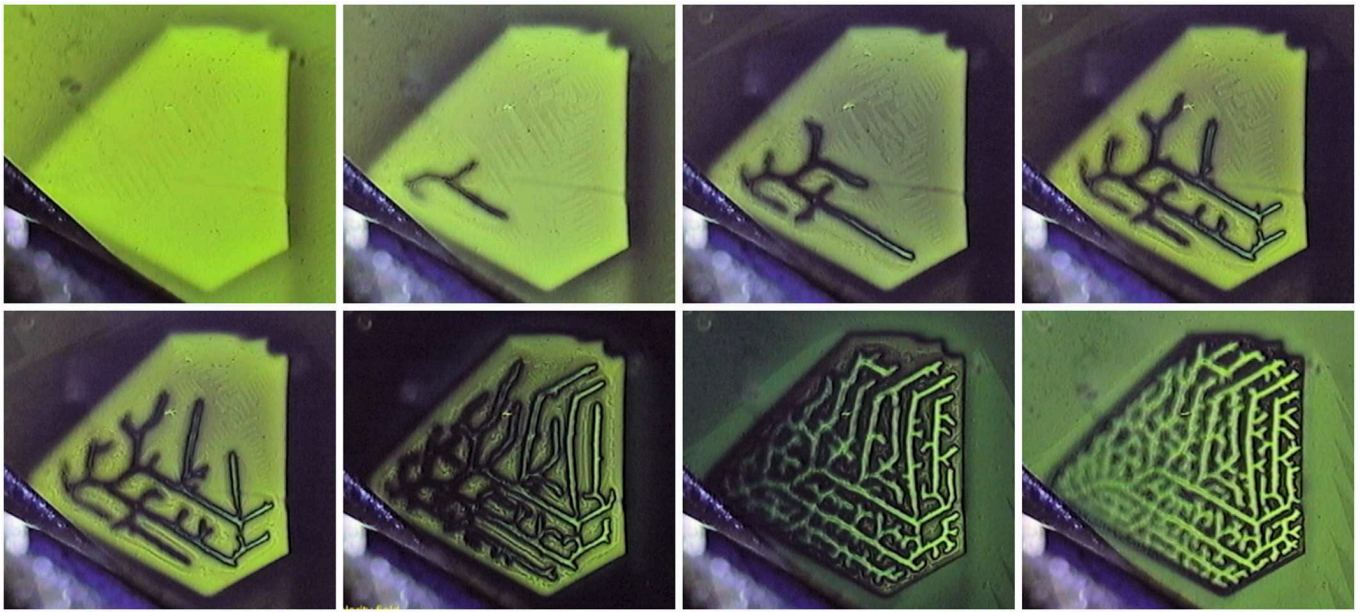}
\caption{(Color online) Turning a magnetic field down after field cooling in 1500 Oe to $T = 50$ K. Top row, left to right: $H=$  1500, 1100, 800, and 500 Oe. Bottom row, left to right: 200, 0, -300, -800 Oe.}
\label{turnoff}
\end{figure*}

\begin{figure}[tb]
\includegraphics[width=9cm]{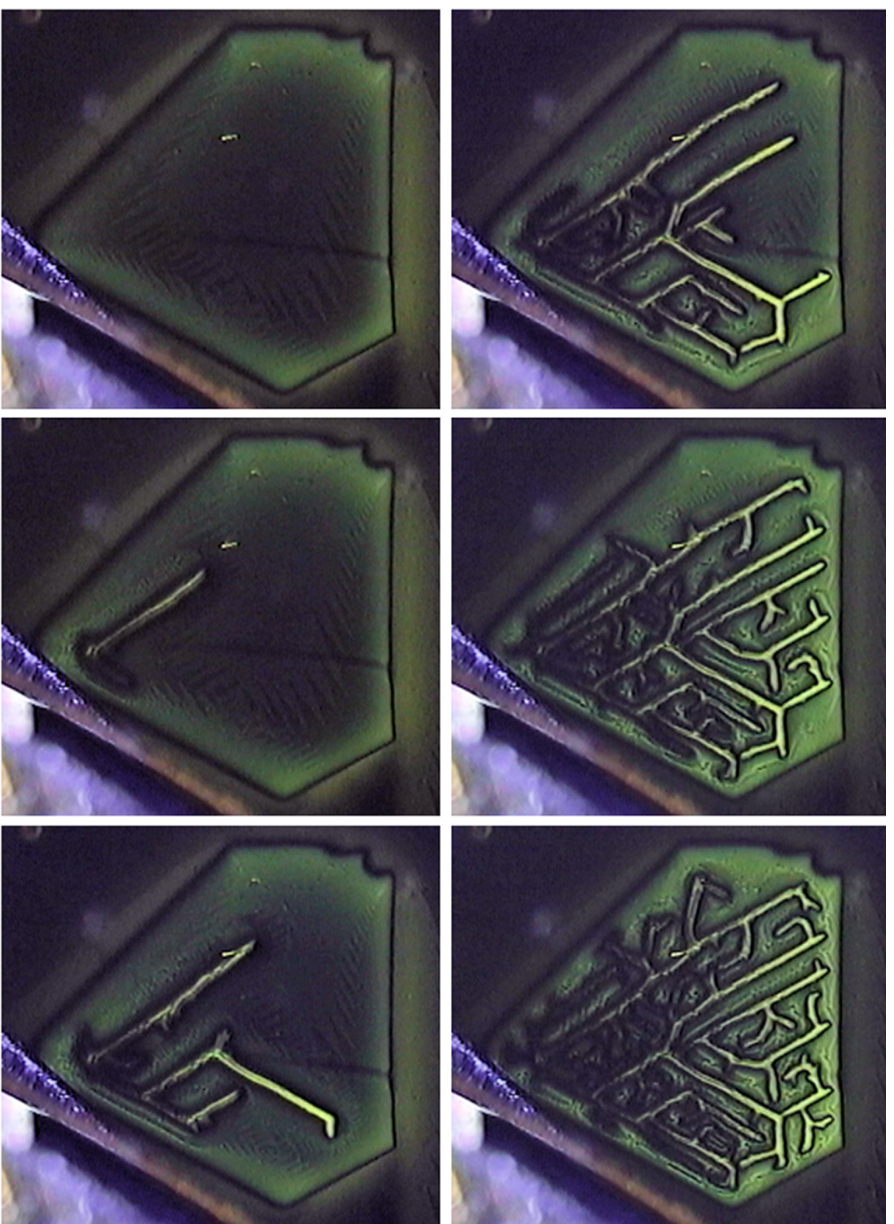}
\caption{(Color online) Warming up after cooling in 1.5 kOe field and turning field off at 5 K. Left column, top to bottom: 5 K, 50 K, 55 K - right column: 60 K, 70 K, 80 K.}
\label{warmup}
\end{figure}

In the present work, a triangular sample of area 0.67 mm$^2$ by 0.05 mm thick was mounted on a sapphire rod and inserted into the tank coil of a TDR mounted on a $^{4}$He cryostat operating at a frequency of 27 MHz.  Temperature dependent $\chi_{rf}$ was measured on warming and cooling in dc bias fields up to 1 kOe.  The dc field was parallel to the ac excitation field, both of which were aligned with the magnetic easy axis of the sample.  In addition to constant field temperature sweeps, $\chi(H)$ data in the range $-2.6$ kOe$\leq H\leq2.6$ kOe were taken at several temperatures to explore the possibility of field dependent effects.

In metallic samples the dynamic susceptibility, $\chi_{rf}$, is composed of both a spin susceptibility and a diamagnetic skin--effect screening. The effect of the latter is determined by the chosen measurement frequency and the sample's resistivity. In the vicinity of a ferromagnetic phase transition the spin susceptibility dominates. Additionally, within the measurement frequency range employed here, domain wall resonance effects may become important (see e.~g.~Ref.~\onlinecite{saitoh-2004}).  This suggests TDR measurements may detect effects associated with domains. The resistivity of \fe~does not show any special features at and below $T_C$.  It merely decreases due to a gradual loss of spin-disorder scattering \cite{fe-morosan-2007}.  Therefore, the skin - effect contribution to the measured susceptibility manifests as a monotonic background.

\section{\label{results}Results and Discussion}
Figure \ref{remag} shows the appearance of ferromagnetic domains after the sample was cooled to 5 K in $H=-1000$ Oe magnetic field and the field was slowly increased.  In these and all other MO images presented herein, the c-axis of the crystal is pointing out of the page.  Field cooling resulted in a monodomain state for the crystal.  In order to nucleate a domain wall, a positive field of $H=$xx Oe was necessary.  The persistence of a monodomain state at 5 K in an oppositely oriented field implies either a small magnetic moment per iron or a large magnetocrystalline anisotropy or both.  The saturated magnetic moment per iron is approximately 4 $\mu_{B}$ which is a moderate magnetic moment.  In conjunction with the previously reported $M$-$H$ data \cite{fe-morosan-2007}, these images suggest a very high anisotropy for this compound.

As the field is further increased, unusual, dendritic domains form.  The domains nucleate at a weak spot, probably due to a local imperfection.  The overall domain structure observed under different conditions is different (compare Figs.~\ref{remag} \ref{turnoff} and \ref{warmup}), so there is no one-to-one correlation between the distribution of the domains and the structural features of the crystal.  However, the gross features of the domain evolution do reflect the general structure of the underlying hexagonal lattice.  This indicates that, in addition to the large c-axis magnetic anisotropy, there exists significant in-plane anisotropy with respect to domain wall motion.  The propagation of a domain dendrite shows that the walls are much more weakly pinned along directions parallel to the crystallographic hexagonal lattice as compared to perpendicular wall motion.

Field cooling to a higher temperature ($T=50$ K) shows that the monodomain state is broken by lowering the applied field without changing the field direction (Fig.~\ref{turnoff}).  Again, the unusual, dendritic domains form in much the same way as at the lower temperature.  Fig.~\ref{warmup} shows how temperature variation affects domain formation.  The sample was cooled to $T=5$ K in an applied field of 1500 Oe.  The field was lowered to zero and the temperature was slowly warmed.  No domains were observed until $T\approx 50$ K, at which temperature the domains grew in a dendritic fashion similar to that observed for the constant temperature field sweeps.  For real time videos, see Ref.~\onlinecite{video}.

It should be emphasized that the observed contrast in the MO images is due to the different domains and not the domain walls themselves.  It is easy to consider such narrow domains as walls.  However, the domain walls separate the regions of different contrast (or color, if read online).  This distinction is most evident in the zero field images of Fig.~\ref{turnoff} bottom row, second from left and all of Fig.~\ref{warmup} where the domain walls show up as fine, dark lines surrounding the dendritic structure.

\begin{figure}[tb]
\includegraphics[width=9cm]{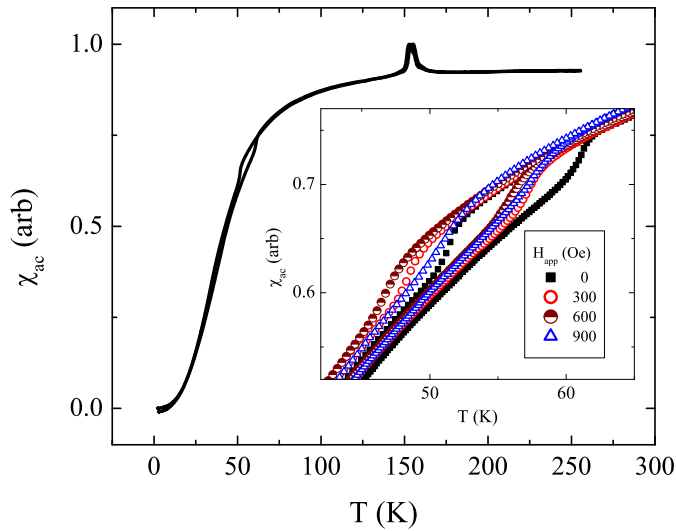}
\caption{(Color online) $\chi_{rf}(T)$ at $H=0$ measured on warming and cooling in \fe~crystal.  \textit{Inset:} Detail of the hysteresis region in applied bias fields.  Both the rf-probe field and the static bias field were aligned with the crystallographic c-axis.}
\label{zerowinset}
\end{figure}

\begin{figure}[tb]
\begin{center}
\includegraphics[width=9cm]{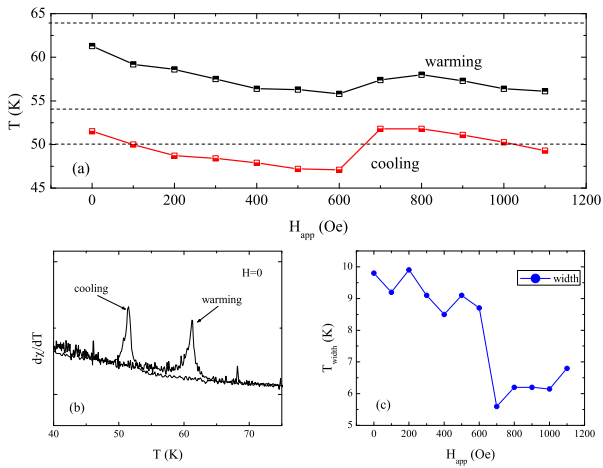}
\caption{(Color online) Breakdown of the hysteresis region in applied fields.  Panel (a) presents the boundaries of the hysteresis vs.~applied field.  The dashed lines are temperatures at which field scans were taken.  Panel (b) is the zero field $d\chi/dT$ vs.~$T$ demonstrating how the boundaries in (a) were determined.  Panel (c) is the width of the hysteresis as defined in the text.}
\label{3way}
\end{center}
\end{figure}

We now turn to a discussion of the dynamic magnetic susceptibility. Figure~\ref{zerowinset} shows $\chi_{rf}(T)$ in arbitrary units for a single crystal of \fe~measured at $H=0$ in the temperature range 2--250 K upon warming and cooling. The ferromagnetic transition is observed as a peak in $\chi_{rf}$ at $T_{C}\approx 155$ K. All rf--$\chi$ data are normalized to 1 at this peak temperature.  Below $T_{C}$ the measured susceptibility drops in accordance with the resistivity decrease \cite{fe-eibschutz-1981}.  In the vicinity of 55 K there is an observed hysteresis characterized by breaks in the slope of $\chi_{rf}$ vs.~$T$.  The inset shows the effect of applied static fields on this hysteresis. Runs performed in field were field cooled in the next lowest field and then $H$ was increased to the value for the run.  To illustrate, the $\chi_{rf}(T)$ data collected in an applied field of 500 Oe was field cooled in 400 Oe.  At base temperature ($\sim$2 K) the field was then increased by 100 Oe at 10 Oe/s.  A plot of $d\chi/dT$ vs.~$T$, as shown in Fig.~\ref{3way}(b), allows for a precise determination of the maximum change in slope for both the warming and cooling curves which are taken to delimit the hysteresis region.  In Fig.~\ref{3way}(a) hysteresis region is marked as a function of field.  Horizontal dashed lines denote temperatures at which field scans were performed.    Figure \ref{3way}(c) shows the width of the hysteresis defined as the difference between the upper and lower temperatures from Fig.~\ref{3way}(a).  First, for low fields the hysteresis shifts to lower temperatures with a gradually decreasing width.  At a critical field of approximately 700 Oe there is a dramatic increase in the low temperature boundary and a corresponding decrease in the width of the hysteresis.  Thereafter, the width of the hysteresis grows slightly.

\begin{figure}[tb]
\begin{center}
\includegraphics[width=9cm]{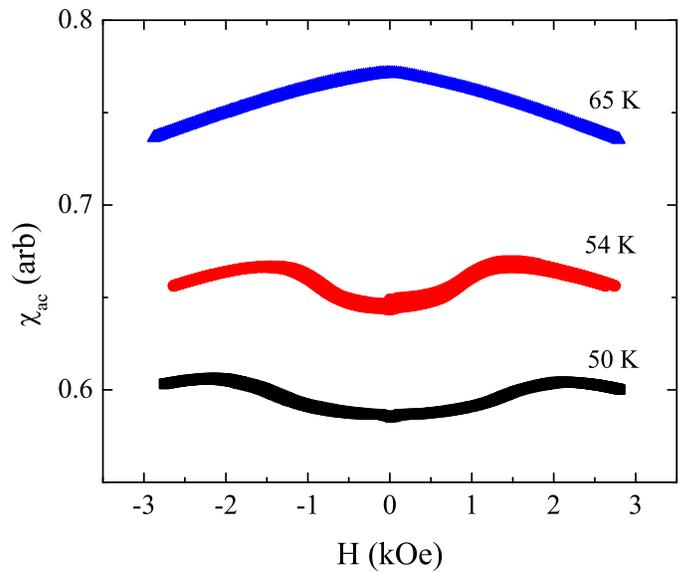}
\caption{$\chi_{rf}$ vs.~$H$ curves for three temperatures near the hysteresis region.  The arbitrary units in this figure are the same as those in Fig.~\ref{zerowinset}.}
\label{chiH}
\end{center}
\end{figure}

\begin{figure}[tb]
\begin{center}
\includegraphics[width=9cm]{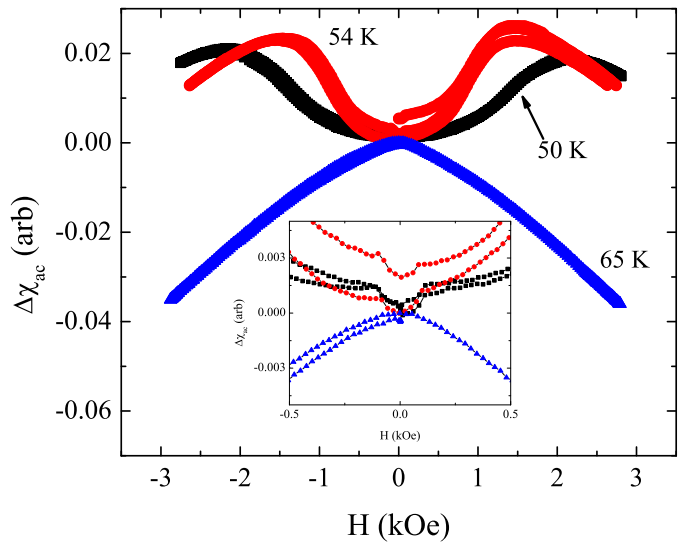}
\caption{$\Delta\chi$ vs.~$H$ for the three field scans.  \textit{Inset:} Detail of the low field region.}
\label{deltachi}
\end{center}
\end{figure}

Figure ~\ref{chiH} presents the field dependent susceptibility data at three temperatures in and above the hysteresis region.  In all curves the sample was zero field cooled, warmed to the measurement temperature and the field was swept up to +2.6 kOe, down to -2.6 kOe and back to zero.  As a check on the polarity of the field, experiments were also run sweeping to negative field first.  A slight zero field difference in $\chi_{rf}$ exists, most notably for the 54 K trace.  However, this is likely due to a small temperature drift on the order of 0.4 K as determined by a comparison of the zero field temperature data.  $\chi_{rf}(T=50$ K,$H)$ data fall below the zero field thermal hysteresis, but at a field of about 100 Oe it crosses the low temperature boundary.  At $T=54$ K the data begins in this hysteresis region, while at $T=64$ K the sweep is done above it.  From Fig~\ref{3way}(a) it is obvious that the lowest temperature field scan crossed the low temperature hysteresis boundary at 100 Oe.  At about 650 and 1000 Oe it crossed the interpolated low temperature boundary twice more.  The mid temperature scan ($T=54$ K) did not cross any hysteresis boundary.  However, the qualitative features in both of the lower temperature scans are similar.  To better compare the three field scans Fig.~\ref{deltachi} shows the change in $\chi_{rf}$ due to field taking the zero field susceptibility as a reference point.  The high temperature field sweep may be understood in the context of polarizing spins.  As the applied field increases there is an increasing preference for the spins to align with the field.  The dynamic response of the moments in the sample decreases as a consequence and the rf susceptibility reflects this.

Within the domain formation temperature window, the effect of a magnetic field is to increase the relative fraction of domains parallel to the field at the expense of all others.  The effect of the rf probe field is to change the alignment of the spins by a small amount.  In a dc bias field there is a preferred direction for the spins to point.  Keeping in mind that the magnetic easy axis is aligned with the dc and rf fields, a spin residing in a domain parallel to the applied field will contribute a greater moment change in response to the rf field than one in an antiparallel domain.  This can be seen by considering the thermal effects.  In general a spin will not point exactly along (parallel or antiparallel) the line of applied field.  Rather it will be tilted away at some angle due to thermal randomization.  During the half cycle when the rf field is parallel with the applied field there will be a slightly greater tendency for all spins to point in that direction. Those spins in domains whose magnetization points along the applied field will rotate more readily toward the field direction as this will minimize both the Zeeman and anisotropy energies.  Spins in domains with antiparallel magnetizations will also tilt toward the applied field thereby reducing Zeeman energy. However, in this case there is an increase in anisotropy energy.  Therefore, the change in the bulk moment from these antiparallel domains is reduced compared with the parallel domains.  On the other half of the rf cycle both parallel and antiparallel spins are working against the Zeeman energy from the bias field and there will not be as great a response.  So as the fraction of sample in domains parallel to the applied field grows the measured $\chi$ grows as well.  At some point, the field has essentially saturated the sample and driven it into a single domain.  Further increasing of the field increases the Zeeman energy effectively 'freezing out' the response of the spins.

The low field region of the 50 and 54 K scans show a characteristic field of 100 Oe observed both upon increasing and decreasing the dc bias field.  Below this field the susceptibility grows rapidly whereas immediately above 100 Oe the change in $\chi_{rf}$ is much more gradual.  In a zero field cooled state, the system is not monodomain.  In terms of the preceding argument, the rapidly increasing $\chi_{rf}$ may be explained by the response of any free or nearly free domain walls to the applied field.  As mentioned previously, this frequency band is suitable for measuring domain wall resonance effects.  It is possible that for this material 27 MHz is close to a natural frequency for the domains, in which case at least some of the observed effects would be frequency dependent.  The only way to be certain of this is to perform a careful study of the susceptibility over a frequency window of $\pm$10--20 MHz.

If Fe$_{1/3}$TaS$_{2}$ is, in fact, itinerant as band structure calculations suggest \cite{fe-dijkstra-1989}, a comparison of the present results with similar data from the other ferromagnetic members of this family may offer clues regarding to how the crossover between local moment and itinerant ferromagnetism proceeds.

In conclusion, unusual ferromagnetic domains were observed in single crystal \fe. Analysis of the rf-magnetic susceptibility and real-time dynamics of the domain formation imply that this material is a local moment ferromagnet with very large magneto-crystalline anisotropy that increases upon cooling.  Further, the evolution of the domain pattern indicates that there is significant anisotropy with respect to domain wall motion within the hexagonal plane.

Work at the Ames Laboratory and Princeton University were supported by the Department of Energy, 
Office of Basic Energy Sciences, under Contracts DE-AC02-07CH11358 and DE-FG02-98ER45706, respectively.

\end{document}